\documentclass{iau}
\usepackage{graphicx}

\title[LOFT: the Large Observatory For X-ray Timing] 
{LOFT: the Large Observatory For X-ray Timing}

\author[Tomaso M. Belloni \& Enrico Bozzo]   
{Tomaso M. Belloni $^1$
 \and  Enrico Bozzo $^2$ (on behalf of the LOFT Consortium)}

\affiliation{$^1$ INAF - Osservatorio Astronomico di Brera, \\ Via E. Bianchi 46, I-23807 Merate,
Italy  \\ email: {\tt tomaso.belloni@brera.inaf.it } \\[\affilskip]
$^2$ISDC, Data Center for Astrophysics of the University of Geneva \\ chemin d'\'Ecogia, 16 1290 Versoix, Switzerland \\ email: {\tt enrico.bozzo@unige.ch}}

\pubyear{2012}
\volume{290}  
\jname{Feeding compact objects: Accretion on all scales}
\editors{C.M. Zhang, T. Belloni, M. M\'endez \& S.N. Zhang, eds.}

\begin{document}

\maketitle

\begin{abstract}
LOFT, the large observatory for X-ray timing, is a new mission concept competing with other four candidates for a launch opportunity in 
2022-2024. 
LOFT will be performing high-time resolution X-ray observations of compact objects, combining for the first time an unprecedented large collecting area 
for X-ray photons and a spectral resolution approaching that of CCD-based X-ray instruments (down to 200~eV FWHM at 6~keV). The operating energy range 
is 2-80~keV. The main science goals of LOFT are the measurement of the neutron stars equation of states 
and the test of General Relativity in the strong field regime. The breakthrough capabilities of the instruments on-board LOFT will permit to 
open also new discovery windows for a wide range of Galactic and extragalactic X-ray sources.  

In this contribution, we provide a general description of the mission concept and summarize its main scientific capabilities. 
\end{abstract}

\firstsection 
\section{Introduction} 
\label{sec:intro} 
\begin{figure}
\begin{center}
 \includegraphics[width=2.6in]{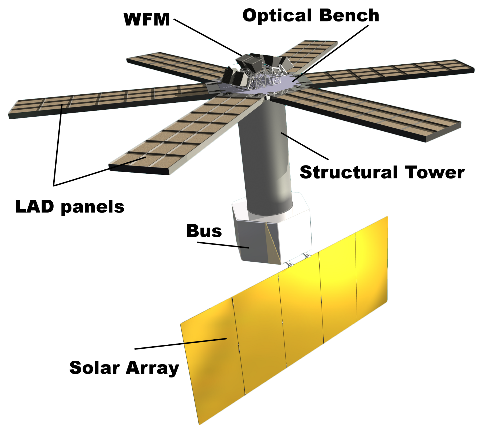}
  \includegraphics[width=2.6in]{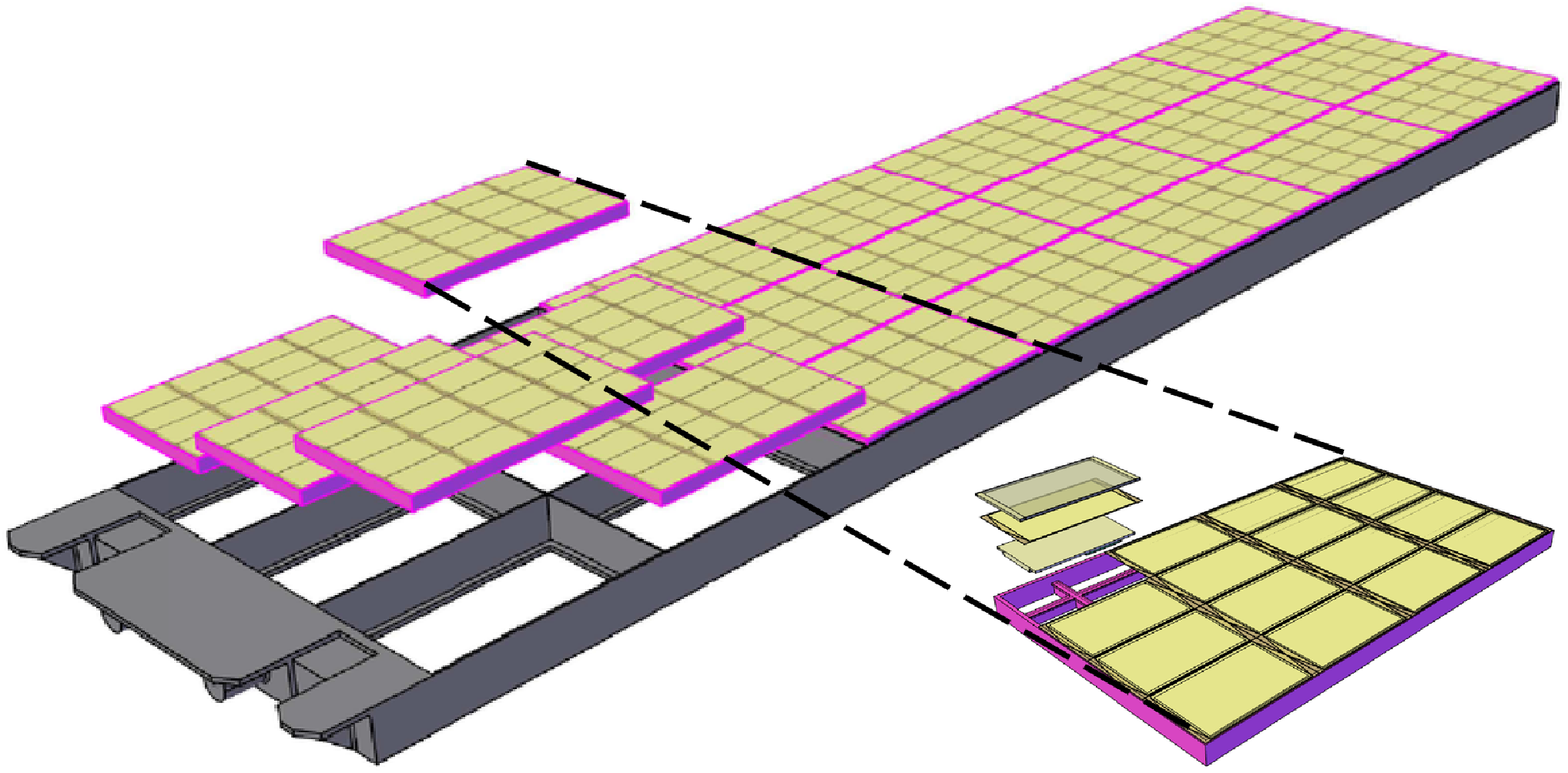}
 \caption{{\it Left}: an artist representation of the LOFT satellite. The LAD instrument comprises 6 panels connected to the optical 
 bench of the spacecraft; the WFM is located on the top of the optical bench. {\it Right}: Breakdown of the LAD instrument. Each panel of the LAD (6 in total) 
 comprises 21 modules and each module is equipped with 16 SDDs.} 
   \label{fig:loft}
\end{center}
\end{figure}

LOFT, the Large Observatory for X-ray Timing, is one of the four mission candidates currently competing for a launch opportunity in 2022-2024. 
The four missions were selected by the European Space Agency (ESA) in February 2011 within the framework of the Cosmic Vision program. 
LOFT was specifically conceived to investigate the behavior of matter in presence of strong gravitational fields and at supranuclear densities 
by performing high-time-resolution X-ray observations of compact objects. Measurements of the rapid X-ray flux and spectral variability 
in these sources can indeed provide constraints on the equation of states of neutron stars (NS) and measurements of black holes masses and spins 
(see Sect.~\ref{sec:science}). A representation of the current view of the LOFT satellite is provided in Fig.~\ref{fig:loft}.  

The Large Area Detector (LAD) is the main instrument on board LOFT. It is a collimated experiment with a field-of-view (FOV) of about 1 degree, reaching  
a peak X-ray sensitive effective area of $\sim$10 m$^2$ at 8~keV (see Fig.~\ref{fig:area}). The expected scientific throughput is of $\sim$240,000 cts/s for a 
source with a flux of 1~Crab in the 2-80~keV energy range. The main operating energy band of the LAD is 2-30~keV, where the expected spectral resolution is 
$<$260~eV (200~eV for single anode events, i.e. about 40\% of the total counts). In the range 30-80~keV, only a coarse energy resolution of $\sim$2~keV 
will be available. The time resolution of the LAD is 10~$\mu$s. 

The second instrument completing the LOFT payload is the Wide Field Monitor (WFM). This is a coded mask imager whose main goal is to detect new transient sources and 
state changes of any X-ray emitter that can be suitable for observations with the LAD. 
The WFM will be observing about 1/3 of the sky at once and will be capable to provide data for each source in this FOV with a spectral and timing resolution 
similar to that of the LAD. 

We provide in Sect.~\ref{sec:instruments} a detailed description of the LOFT payload, and describe in Sect.~\ref{sec:science} some of the most recent improvements 
in the instruments design. All images and simulations  used in the following sections are provided by the LOFT teams\footnote{See 
http://www.isdc.unige.ch/loft/index.php/preliminar-response-files-and-simulated- background and 
http://www.isdc.unige.ch/ loft/index.php/instruments-on-board-loft.}.

\section{LOFT Spacecraft and Instrumentation}
\label{sec:instruments} 

In the current configuration of the LOFT mission (see Fig.~\ref{fig:loft}), 
the LAD instrument comprises six panels attached to the spacecraft optical bench. On the top of it, the five units (10 cameras) of the 
WFM are placed in a way to maximize the sky coverage (\cite[Brandt et al. 2012]{brandt12}).   
\begin{figure}
\begin{center}
 \includegraphics[width=2.6in]{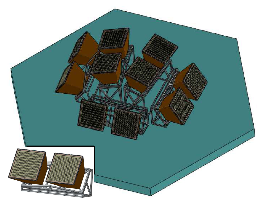}
    \includegraphics[width=2.6in]{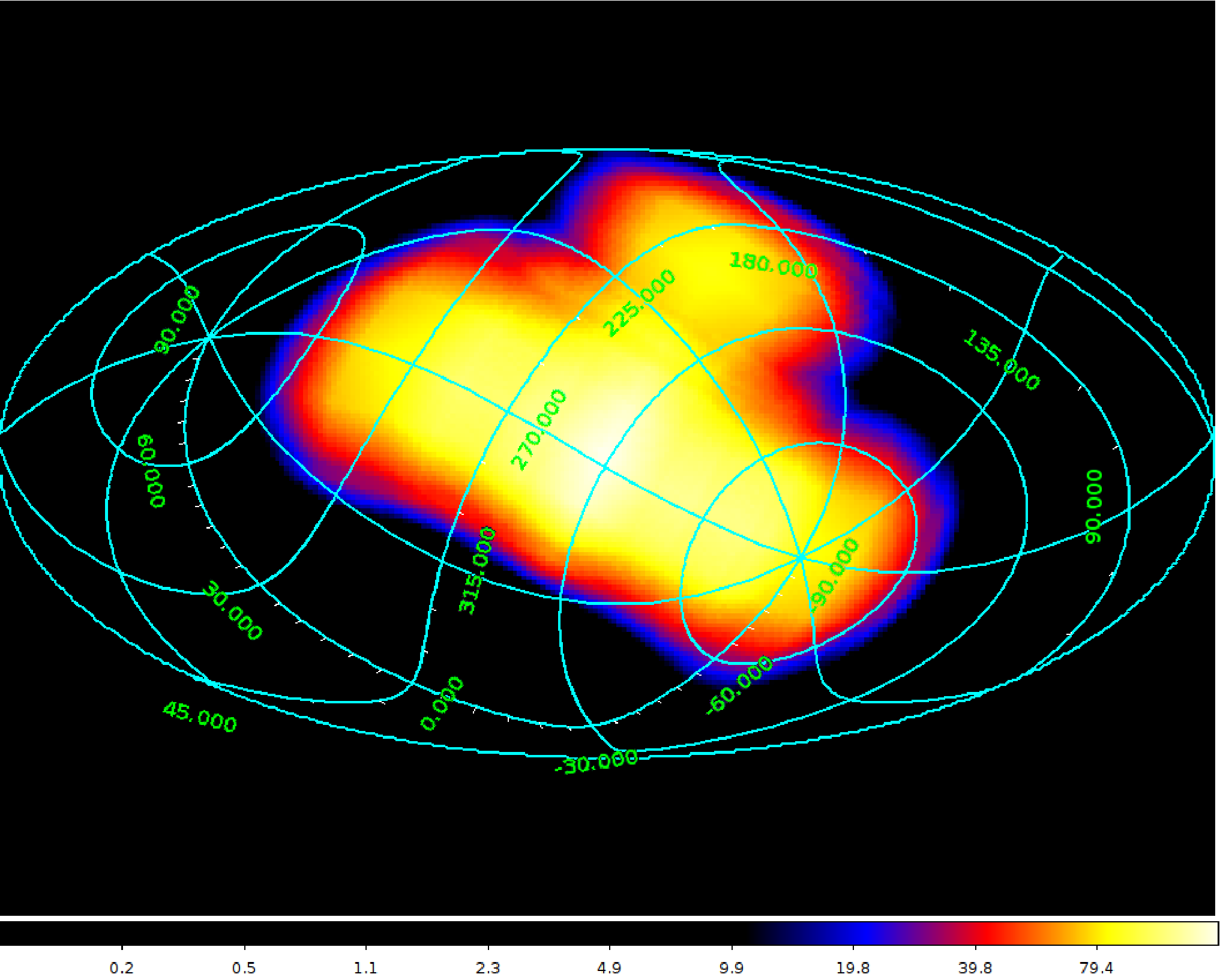}
 \caption{{\it Left}: detailed view of the WFM located at the top of the optical bench. The insert shows the concept of a WFM unit comprising two orthogonal cameras 
 to achieve a full 2D imaging capable system. {Right}: The WFM FOV expected when the instrument points toward the Galactic center. The color bar at the bottom 
 gives the effective area in units of cm$^{2}$.}  
   \label{fig:wfm}
\end{center}
\end{figure}

\begin{figure}
\begin{center}
 \includegraphics[width=2.6in]{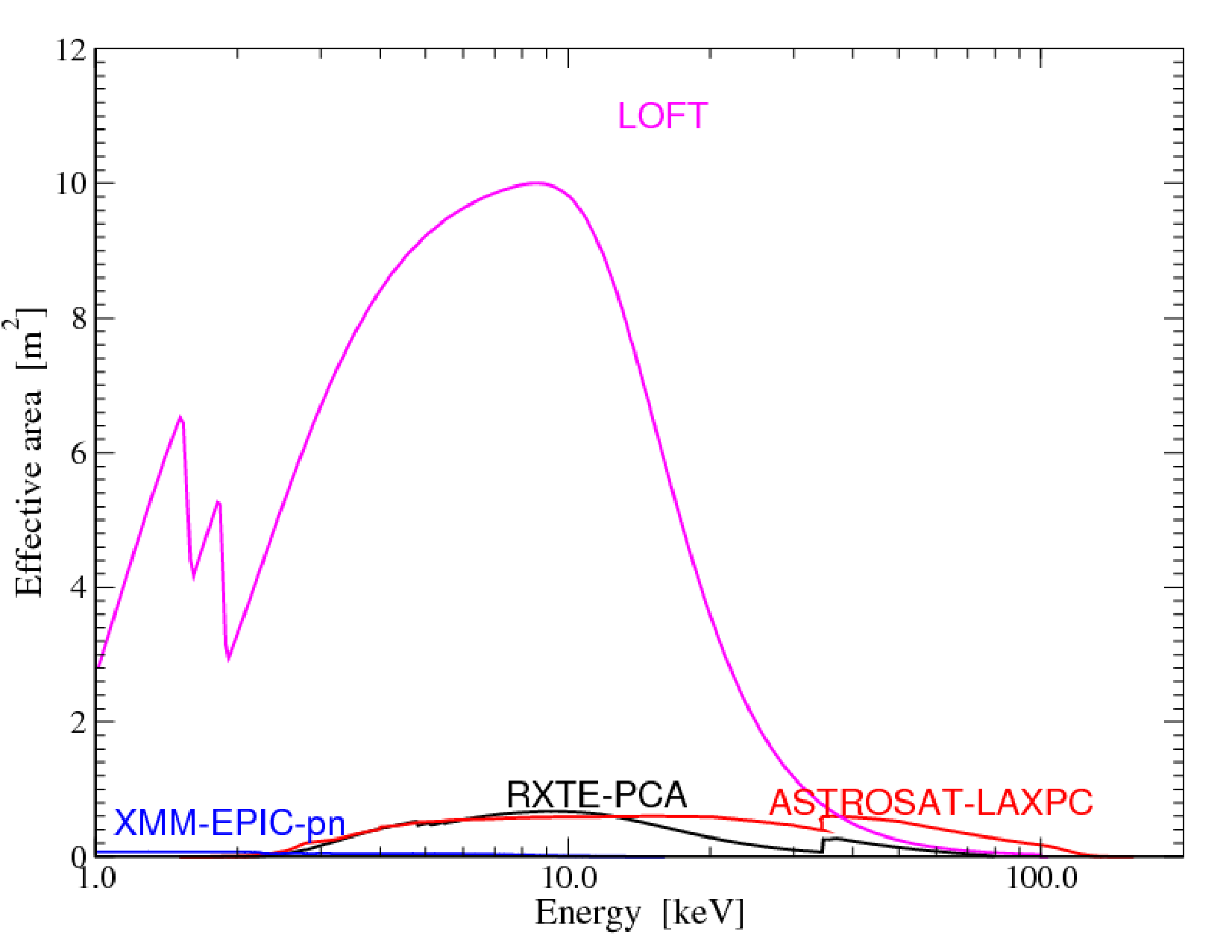}
 \includegraphics[width=1.2in]{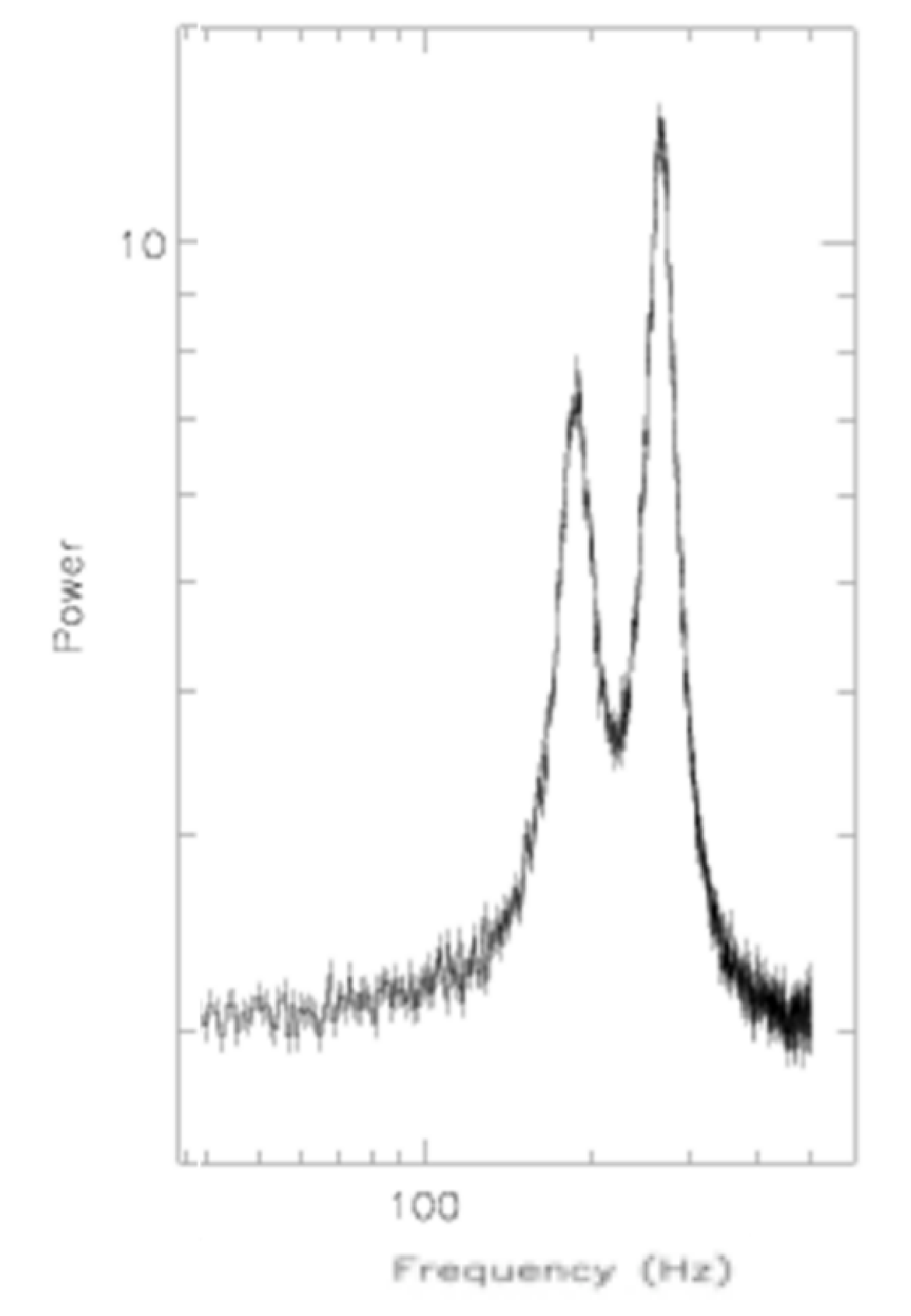}
 \caption{{\it Left}: Plot of the LAD effective area as a function of the energy according to the instrument requirements (see Table~\ref{tab:lad}). The effective area of 
 some other instruments is reported for comparison. {\it Right}: A simulated power-spectrum of the black hole candidate XTE J1550-564 as observed by RXTE/PCA 
 (\cite[Miller et al. 2001]{miller01}). For the simulation it is assumed that the two QPOs have frequencies of $\nu_1$=188~Hz, $\nu_2$=268~Hz and the fractional 
 rms are 2.8\% and 6.2\%, respectively. The source flux is 1~Crab (3-20~keV). The exposure time used for the LAD simulation is 1~ks. The improvement in the S/N of this 
 detection should be compared by looking at Fig.~4 in \cite[Miller et al. (2001)]{miller01}, where the original RXTE/PCA exposure time of 54~ks permitted only a detection at 
 3-4~$\sigma$ confidence level.} 
   \label{fig:area}
\end{center}
\end{figure}

The LAD is the prime instrument onboard LOFT. Its geometric area is $\sim$15~m$^2$ and provides a peak effective 
area for X-ray observations that is a factor of 20 larger than that of any previously flown X-ray experiment (see left panel of Fig.~\ref{fig:area}). 
The LAD achieves an energy resolution (FWHM) better than 260~eV at 6~keV (end-of-life, EoL). The two enabling technologies of the LAD are the 
large area Silicon drift detectors developed originally for the ALICE experiment at CERN (\cite[Vacchi et al. 1991]{vacchi91}; 
\cite[Rashevski et al. 2002]{Rashevski02}) and the capillary plates collimators designed for the ESA BepiColombo mission 
(\cite[Fraser et al. 2010]{fraser10}). 
The SDDs have been optimized for LOFT to reach a configuration in which each detector is now 450~$\mu$m thick and provides a total active area of 
76~cm$^2$, read-out by two rows of anodes with a pitch of 970~$\mu$m and a drift channel of 35~mm. 
The capillary-plate collimator was chosen to make advantage of the lead glass with millions of micro-pores, already proven to be efficient in collimating 
X-ray photons within a field of view (FOV) of $\sim$1~degree and energies up to 30-40~keV.  
In the current LOFT design, each of the SDD is equipped with a single collimator tile ($\sim$80 cm$^2$ in area). There are in total 
6 Detector Panels, each hosting 21 Detector Modules. 16 SDDs are allocated within each module (see also \cite[Zane et al. 2012]{zane12}). 
A total of 2016 SDDs are needed for the complete LAD instrument. 
The 6 LAD panels will be initially stowed in the launcher, and then deployed once in space 
(\cite[Feroci et al. 2011]{feroci11}). A summary of the LAD capabilities is provided in Table~\ref{tab:lad}.  

The WFM is a wide field coded mask imager working mainly in the 2-50~keV energy range (see Fig.~\ref{fig:wfm}). 
The instrument is designed from the heritage of the SuperAGILE experiment (\cite[Feroci et al. 2007]{feroci07}), but presents 
very notable improvements in the low energy threshold, energy resolution and imaging capabilities. These are provided by the usage of similar 
SDDs to that employed for the LAD but optimized for imaging purposes (\cite[Campana et al. 2011]{campana11}). 
Due to this optimization, the time resolution is kept of the order of $\sim$10$\mu$s, whereas the spectral resolution is somewhat 
lower than that achieved by the LAD (300~eV FWHM at 6~keV). The WFM SDDs are able to measure the impact point of a photon with an accuracy that 
corresponds to few arcmin along the anode direction and few degrees along the drift direction (\cite[Evangelista et al. (2012)]{evangelista12}). 
For this reason, each of the WFM unit (5 in total, see Fig.~\ref{fig:wfm}) comprises two orthogonal cameras to achieve a full 
2D imaging capability (\cite[Brandt et al. (2012)]{brandt12}). This design has also the advantage of providing a strong redundancy 
in case of failure of one camera. In that case, indeed, a unit will be left with a fine resolution in one direction and a coarse resolution 
in the other, still ensuring some coverage of the sky in the corresponding region). 
The imaging properties of the WFM are extensively discussed in \cite[Donnarumma et al. 2012]{donnarumma12}.
In Fig.~\ref{fig:wfm} we show as an example the WFM FOV during a pointing toward the Galactic Center. 
The WFM will cover more than 1/3 of the sky at once, corresponding to 50\% of the sky accessible to the LAD at any time. 
A summary of the WFM capabilities is provided in Table~\ref{tab:lad}.  
\begin{table}
\begin{minipage}[b]{0.46\linewidth}
\tiny
\begin{tabular}{@{}lll@{}}
\multicolumn{3}{c}{\bf LAD} \\
\hline
Parameter & Requirement & Goal \\
\hline
Energy range & 2--30~keV (nom.) & 1--30 keV (nom.) \\
             & 30--80~keV (exp.)& 1--80 keV (exp.) \\
\vspace{-0.2cm}\\
Eff. area & 10.0~m$^2$ (8~keV) & 12~m$^2$ (8~keV) \\
          & 1.0~m$^2$ (30 keV) & 1.2~m$^2$ (30 keV) \\
\vspace{-0.2cm}\\
$\Delta$E & $<$260~eV & $<$200~eV \\
(FWHM, @6 keV)  & (200~eV, 40\%)$^a$  & (160~eV, 40\%)$^a$ \\
\vspace{-0.2cm}\\
FoV (FWHM) & $<$60 arcmin & $<$30 arcmin \\
\vspace{-0.2cm}\\
Time res. & 10 $\mu$s & 7 $\mu$s \\
\vspace{-0.2cm}\\
Dead time & $<$1\% (@1 Crab) & $<$ 0.5\% (@1 Crab) \\
\vspace{-0.2cm}\\
Background flux & $<$10 mCrab & $<$ 5 mCrab \\
\vspace{-0.2cm}\\
Max. flux (steady) & $>$0.5 Crab &  $>$0.5 Crab \\
\vspace{-0.2cm}\\
Max. flux (peak) & $>$15 Crab &  $>$30 Crab \\
\hline
\multicolumn{3}{l}{$a$: Refers to single-anode events.} \\
\end{tabular}
\end{minipage}
\tiny
\centering
\hspace{0.5cm}
\begin{minipage}[b]{0.46\linewidth}
\begin{tabular}{@{}lll@{}}
\multicolumn{3}{c}{\bf WFM} \\
\hline
Parameter & Requirement & Goal \\
\hline
Energy range & 2--30~keV (nominal) & 1--30 keV (nom.) \\
             & 30--80~keV$^{a}$ & 1--80 keV$^{a}$ \\
\vspace{-0.2cm}\\
$\Delta$E (FWHM)& $<$500 eV & $<$300 eV \\
\vspace{-0.2cm}\\
FoV (FWHM) & 1 pi steradian &  1.5 pi steradian\\
& around the LAD & around the LAD\\
& pointing & pointing including\\
&  & anti-Sun direction\\
\vspace{-0.2cm}\\
Angular res. & 5 arcmin & 3 arcmin \\
\vspace{-0.2cm}\\
Position Accuracy& 1 arcmin & 0.5 arcmin \\
\vspace{-0.2cm}\\
Sensitivity & 5 mCrab & 2 mCrab \\
(5$\sigma$, 50 ks)  \\
\vspace{-0.2cm}\\
Sensitivity  & 1 Crab & 0.2 Crab \\
(5$\sigma$, 1 s) \\
\hline
\multicolumn{3}{l}{$a$: Events in the 30-80~keV energy range are used} \\ 
\multicolumn{3}{l}{to monitor contamination from bright sources.} \\
\end{tabular}
\end{minipage}
\label{tab:lad}
\caption{\scriptsize {Scientific requirements and goals of the LAD and WFM.}}
\end{table}

\subsection{Latest mission improvements}
\label{sec:improvements} 

Due to its large field of view, wide energy coverage and imaging capabilities, the WFM is 
expected to detect and localize a large number of gamma-ray bursts and other fast transients every year (\cite[Brandt et al. 2012]{brandt12}). 
The orientation of one of the WFM unit in the anti-Sun direction makes also particularly favorable the follow-up of any interesting event detected by this instrument 
with ground-based facility that will be operating at the time of LOFT. In order to optimize the synergy between LOFT and these facilities, 
the WFM onboard data processing is endowed with a triggering and imaging system to calculate the coordinates
of the transient event in real-time on-board (LBAS, i.e. LOFT Burst Alert System). This information is broadcast to the 
ground through a VHF transmission system to within $\lesssim$30~s from the detection.  
Due to possible telemetry limitations, the maximum sustainable rate of triggers might be of one per orbit (i.e. one every 90~min.). 

Some effort has been recently devoted to increase the fraction of the sky accessible to the LAD at any time with respect to 
that defined in the original LOFT configuration\footnote{See http://sci.esa.int/science-e/www/object/index.cfm?fobjectid=49448}.   
This will improve the flexibility of the mission in response to requests for target of opportunity observations (ToOs). However, pointings performed 
outside the nominal sky region might be affected by a somewhat degraded spectral resolution (a factor 1.5 worse than the
nominal value) due to the more challenging thermal environment at which the SDDs have to be operated.  
This is, however, not expected to be an issue for all those observations that do not deal with narrow spectral features and do not require the maximum achievable spectral 
resolution.    

\section{LOFT Science}
\label{sec:science} 

The main science goals of LOFT are the study of matter in presence of extreme densities and strong gravitational 
fields. Through its unique timing capabilities and fine spectral resolution, LOFT will be able to 
provide unprecedented constraints to the neutron star equation of state (EoS) and test 
general relativistic effects that are expected to affect X-rays emitted to within a few gravitational radii ($r_{\rm g}$) 
from a central black hole (so far General Relativity has been tested only in the weak-field regime, i.e. for 
$r_{\rm g}$$\sim$10$^5$-10$^6$). 
Beside this, the unique capabilities of the LAD and WFM will also contribute to dramatically deepen our understanding of the physics 
of a wide range of Galactic and extra-galactic sources. The two instruments 
will permit spectroscopic and variability studies down to previously unexplored small timescales (few $\mu$s), and provide prompt fine 
spectral and timing resolution data for any impulsive bright events originating from Galactic sources as well as cosmic gamma-ray bursts. 
About 50\% of the total LOFT observational time will be devoted to this ``observatory science''.  

\subsection{Main Science Goals}
\label{sec:mainscience} 

Obtaining sufficiently accurate (within a few \%) measurements of neutron stars masses and radii is one of the most direct 
probes we have to test our currently understanding of matter at supranuclear densities, as this can constraint the different 
EoS models proposed for these objects. LOFT will be able to measure the neutron star mass and radius down to an accuracy of 
a few percent independently by using different techniques (e.g., pulse profile fitting in millisecond X-ray pulsars 
and spectral fitting of the type-I X-ray bursts observed from neutron star low mass X-ray binaries; \cite[Feroci et al. 2011]{feroci11}). 
The top-level science goals of LOFT related to the determination of dense matter 
EOS are summarized in \cite[Feroci et al. (2012)]{feroci12}.

\begin{figure}
\begin{center}
 \includegraphics[width=4.6in]{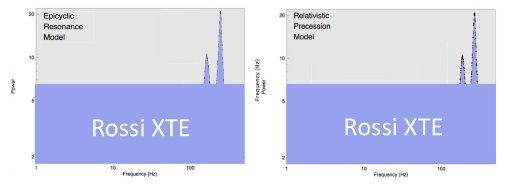}
 \includegraphics[width=4.6in]{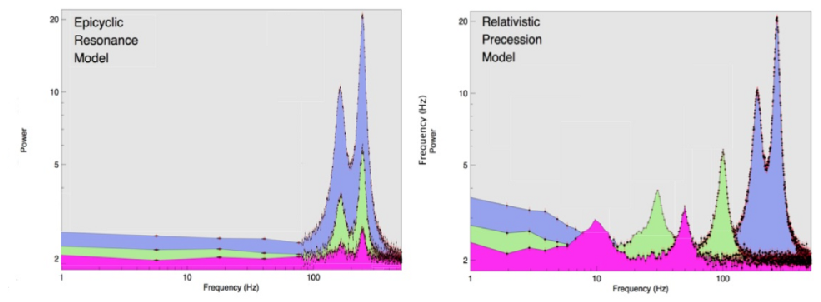}
 \caption{The goal of the LAD is not only to detect QPOs with higher S/N, as shown in Fig.~\ref{fig:area} (right panel). In this figure we show that the LAD will 
 also be able to follow the evolution in time of the QPOs, removing the observational constraints of its predecessor the RXTE/PCA 
 (represented with a blue obscuring rectangular in the upper panel). In the bottom two panels we show a simulation in which the same QPO in Fig.~\ref{fig:area} is evolved 
 following the prediction of the epicyclic resonant model (\cite[Abramowicz et al. 2001]{abramowicz01}) and the relativistic precession model (\cite[Stella et al. 1999]{stella99}). In the simulations we assumed:  
 flux of 1~Crab and fractional rms 2.8\% and 6.2\% for the blue QPOs, flux of 400~mCrab and fractional rms 1.4\% and 3.1\% for the green QPOs, flux of 300 mCrab and 
 fractional rms 0.7\% and 1.5\% for the magenta QPOs. The exposure time is 16~ks in all cases.}
   \label{fig:qpoevo}
\end{center}
\end{figure}

\begin{figure}
\begin{center}
 \includegraphics[width=3.8in]{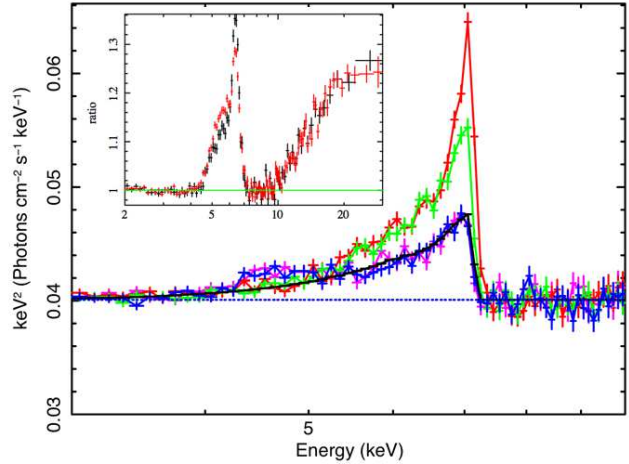}
 \caption{Simulation of the detection with the LAD of the iron line in four different orbital phases of the accretion flow in 
 the AGN MGC\,6-30. The simulation assumes an average flux of 3~mCrab and an exposure time of 16~ks for each of the orbital phases 
 (each phase is observed twice; courtesy of A. de Rosa).}
   \label{fig:line}
\end{center}
\end{figure}

By making advantage of its unique throughput, LOFT will also be able to test General Relativistic effects produced by the strong gravitational field 
of a black hole onto accreting matter orbiting close-by (a few $r_{\rm g}$). One of the most promising way to carry out such investigations is to 
observe quasi-periodic oscillations (QPOs) arising from the X-ray emission of the accretion flow. The dynamical time-scales of the inner accretion flow 
are typically of few milliseconds, and thus the interpretation of the highest frequency QPOs involves fundamental frequencies of the motion of matter
orbiting in disk regions dominated by the strong black hole gravitational field (e.g., the relativistic radial and vertical epicyclic frequencies or the 
relativistic nodal and periastron precession).  
Distinguishing among these possibilities have been proved impossible so far, due to the limited observation capabilities of the previous generation of X-ray 
timing instruments. The enhanced capabilities of the LAD will permit to dramatically increase the S/N of any of the QPO feature (see right panel 
of Fig.~\ref{fig:area}), allowing to measure QPOs accurately within their coherence time and follow their evolution in time with the source X-ray flux 
(see Fig.~\ref{fig:qpoevo}).  
This will secure access to still untested General relativistic effects, such as frame-dragging, strong-field periastron precession, and the
existence of an innermost stable orbit around black holes. 
An independent investigation of the motion of the accretion flow orbiting close to the central black hole can be carried out through the 
observation of the fluorescence Fe K line profile emitted by this material at different orbital phases, by making advantages of the fine 
resolution of the LAD. Variation in the line profile 
are expected as a consequence of the Lense-Thirring precession of the inner disk (at $\sim$r$_{\rm g}$), and provide a tool to measure 
mass and spin of Galactic and extra-galactic black holes. In Fig.~\ref{fig:line}, we show a simulation of such a study in the Active Galactic Nuclei (AGN)  
MGC\,6-30. A summary of the top-level science goals of LOFT with respect to the measure of General Relativistic effects can be found in 
\cite[Feroci et al. (2012)]{feroci12}. 

\subsection{Observatory Science}
\label{sec:observatory} 

\begin{figure}
\begin{center}
 \includegraphics[width=2.5in]{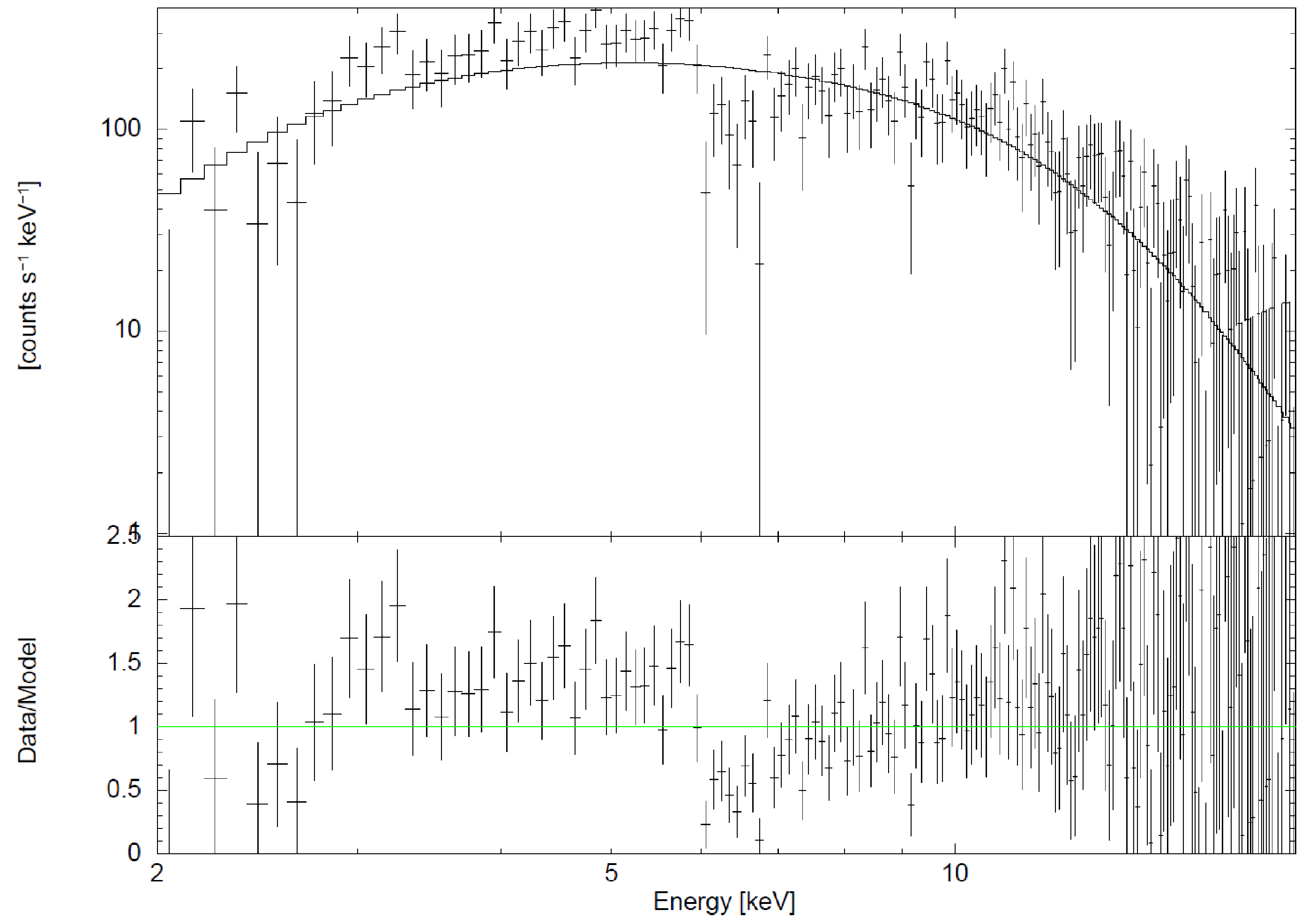}
 \includegraphics[width=2.5in]{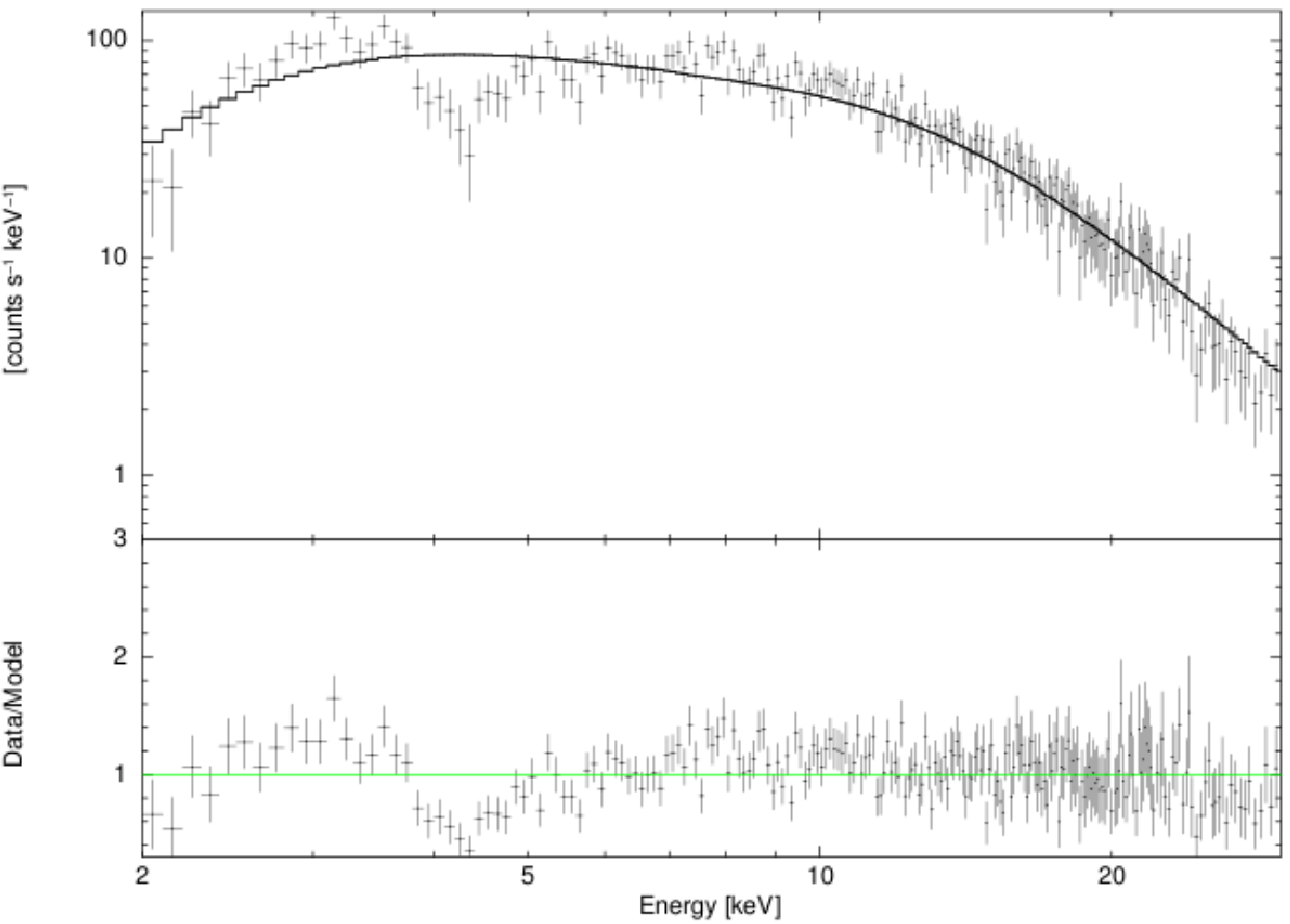}
 \caption{{\it Left}: Absorption edges in energetic type-I X-ray bursts were predicted theoretically by \cite[Weinberg et al. (2006)]{weimberg06} and 
 the only observational evidence of them was reported so far by \cite[in't Zand et al. (2010)]{zand10}. Following their results we simulated here 
 an X-ray burst with a flux of 5~Crab with the WFM including a $\tau$=1 absorption edge at 6~keV. The edge is clearly detected already with 1~s exposure 
 (courtesy of WFM-SIM group). 
 {\it Right}: The only example in the literature of narrow spectral features in the prompt emission of a GRB is reported by \cite[Amati et al. (2000)]{amati00}. Here 
 we simulated an observation of that GRB with the WFM using an exposure time of 13~s. The edge at 3.8~keV is clearly detected (courtesy of WFM-SIM group).}
   \label{fig:edges}
\end{center}
\end{figure}

About 50\% of the total LOFT observational time will be made available for observatory science. 
The breakthrough capabilities of the LAD will revolutionize X-ray observations of any relatively bright  
source down to a limiting flux of few mCrab. Providing an overview of the entire potential discovery space of the LAD 
is not possible in this relatively short contribution. Instead we provide here  
some more examples of the capabilities of the WFM for the observatory science. 
This instrument combines a very wide FOV, covering more than 1/3 of the sky at once, together with a fine timing 
(few $\mu$s) and spectral resolution ($<$300~eV). Through the LBAS system, the position and trigger time of bright events  
can be broadcast to the ground within a delay of $\lesssim$30~s, allowing for quick follow-up with  
ground-based facilities. Due to the large FOV, about 150 GRB and 
thousands type-I X-ray bursts are expected to be detected with the WFM per year. The low energy threshold 
and fine spectral resolution of the instrument will allow to investigate in detail any spectral feature expected during these events, 
including absorption edges. We show two specific simulations in Fig.~\ref{fig:edges}. 
As for the LAD, the range of scientific investigations available to the WFM cannot be sufficiently summarized here. 
As an example, we refer the reader to \cite[Romano et al. (2012)]{romano12} and \cite[Ferrigno et al. (2012)]{ferrigno12} for a description 
of the WFM contribution to the improvement in our knowledge of the Be X-ray binaries and Supergiant Fast X-ray Transients.  

\end{document}